\documentclass[12pt]{article}

\usepackage{amsmath}
\usepackage{amsthm}
\usepackage{amssymb}
\usepackage{color}

\newenvironment{jglist}
{\begin{list}{$\bullet$} {
\setlength{\listparindent}{\parindent}\setlength{\parsep}{0 em} }}
{\end{list}}

\def\innerprod(#1,#2){{\left<#1\,,\,#2\right>}}
\def\Set#1{{\left\{#1\right\}}}

\def\qquadand{\qquad\text{and}\qquad}
\def\quadand{\quad\text{and}\quad}

\def\iv{i_V}
\def\iu{i_U}
\def\ia{i_a}
\def\ib{i_b}
\def\is{i_s}

\def\dual#1{{\widetilde{#1}}}
\def\dualv{{\dual{V}}}
\def\dualu{{\dual{U}}}

\def\deltae{{\dot e}}

\def\man{{M}}

\def\Id{\textup{Id}_4}

\newcommand{\GX}{{\cal G}_X}

\newcommand{\T}[1]{\tau^{#1}}

\newcommand{\gt}{{g_t}}

\newcommand{\PPi}{{\cal P}}
\newcommand{\wdd}{\wedge}

\newcommand{\STAR}{\star\,}

\def\ivF{\iv F}
\def\ivstarF{\iv{{\star}}F}
\def\iuF{\iu F}
\def\iustarF{\iu{{\star}}F}
\def\ivG{\iv G}
\def\ivstarG{\iv{{\star}}G}

\newcommand{\bfe}{{\mathbf e}}
\newcommand{\bfb}{{\mathbf b}}
\newcommand{\bfd}{{\mathbf d}}
\newcommand{\bfh}{{\mathbf h}}
\newcommand{\bfm}{{\mathbf m}}
\newcommand{\bfp}{{\mathbf p}}
\def\Me{{\mathbf e}}
\def\Mb{{\mathbf b}}
\def\Md{{\mathbf d}}
\def\Mh{{\mathbf h}}

\def\PermDE{{\zeta^{\textup{de}}}}
\def\PermDB{{\zeta^{\textup{db}}}}
\def\PermHE{{\zeta^{\textup{he}}}}
\def\PermHB{{\zeta^{\textup{hb}}}}
\def\PermGen{{\zeta^{}}}
\def\PermTen{{Z}}

\def\PermDEO{{\zeta^{\textup{de}}_0}}
\def\PermDBO{{\zeta^{\textup{db}}_0}}
\def\PermHEO{{\zeta^{\textup{he}}_0}}
\def\PermHBO{{\zeta^{\textup{hb}}_0}}
\def\PermGenO{{\zeta^{}_0}}
\def\PermTenO{{Z_0}}
\def\starO{{\star_0}}
\def\daggerO{{\dagger_0}}
\def\gO{g_0}

\def\vO{V_0}
\def\dualvO{{\tilde V}_0}
\def\piO{\pi_0}

\def\PermDET{{\zeta^{\textup{de}}_t}}
\def\PermDBT{{\zeta^{\textup{db}}_t}}
\def\PermHET{{\zeta^{\textup{he}}_t}}
\def\PermHBT{{\zeta^{\textup{hb}}_t}}
\def\PermHBT{{\zeta^{\textup{hb}}_t}}
\def\PermGenT{{\zeta^{}_t}}
\def\PermTenT{{Z_t}}
\def\starT{{\star_t}}
\def\daggerT{{\dagger_t}}
\def\gT{g_t}

\def\vT{V_t}
\def\dualvT{{\tilde V}_t}
\def\piT{\pi_t}
\def\piV{\pi_V}
\def\ivT{i_{V_t}}
\def\LagT{\Lambda_t}
\def\MeT{\Me_t}
\def\MbT{\Mb_t}
\def\MdT{\Md_t}
\def\MhT{\Mh_t}
\def\dualMeT{\dual{\Me}_t}
\def\dualMbT{\dual{\Mb}_t}

\def\PermTenD{{\dot{Z}}}
\def\starD{{\dot\star}}
\def\gD{\dot{g}}

\def\vD{\dot{V}}
\def\dualvD{\dot{\tilde V}}

\def\ivD{i_{\dot{V}}}
\def\LagD{\dot{\Lambda}}

\def\GamTM{{\Gamma T\man}}
\def\GamLamM#1{{\Gamma \Lambda^{#1}\man}}

\def\GenPoyn{{\mathbf s}}
\def\SEM{ stress-energy-momentum }

\def\EM{ electromagnetic }

\def\KK{{\cal K}}
\def\VV{Y}

\begin{document}
\begin{titlepage}
\title{New Perspectives On the Relevance of Gravitation for the Covariant Description of
Electromagnetically
   Polarizable Media}
\author{ T. Dereli\footnote{E.mail: tdereli@ku.edu.tr}
\\{\small Department of Physics, Ko\c{c} University}
\\{\small 34450 Sar{\i}yer,  \.{I}stanbul, Turkey}\\
\\J.  Gratus\footnote{E.mail: j.gratus@lancaster.ac.uk} \\
{\small Department of Physics, Lancaster University and the Cockcroft Institute}\\
{\small Lancaster LA1 4YB, UK},\\
\\R. W. Tucker\footnote{E.mail: r.tucker@lancaster.ac.uk} \\
{\small Department of Physics, Lancaster University and the
Cockcroft Institute}\\ {\small Lancaster LA1 4YB, UK}}

\maketitle


\begin{abstract}
\noindent

By recognising that \SEM tensors are fundamentally related to
gravitation in spacetime  it is argued that the classical
electromagnetic properties of a simple polarisable medium may be
parameterised in terms of a constitutive tensor whose properties
can in principle be determined by experiments in non-inertial
(accelerating) frames and in the presence of weak but variable
gravitational fields. After establishing some geometric notation,
discussion is given to basic concepts of stress, energy and
momentum {\it in the vacuum} where the useful notion of a {\it
drive} form is introduced in order to associate the conservation
of currents involving the flux of energy, momentum and angular
momentum with spacetime isometries. The definition of the \SEM
tensor is discussed with particular reference to its symmetry
based on its role as a source of relativistic gravitation. General
constitutive properties of {\it material continua}  are formulated
in terms of spacetime tensors including those that describe {\it
magneto-electric phenomena} in moving media. This leads to a
formulation of a self-adjoint constitutive tensor describing, in
general, inhomogeneous, anisotropic, magneto-electric bulk matter
in arbitrary motion. The question of an invariant characterisation
of intrinsically  magneto-electric media is explored. An action
principle is established to generate the phenomenological Maxwell
system and the use of variational derivatives  to calculate \SEM
tensors is discussed in some detail. The relation of this result
to tensors proposed by Abraham and others is discussed in the
concluding section where the relevance of the whole approach to
experiments on matter in non-inertial environments with variable
gravitational and electromagnetic fields is stressed.

\bigskip


Maxwell's equations.
\end{abstract}
\end{titlepage}


\section{Introduction}
\label{ch_intro}

The laws of quantum-electrodynamics have been devised to describe
the electromagnetic interactions with matter according to the
tenets of relativistic quantum field theory. However Maxwell's
classical equations remain mandatory for the description of a vast
amount of natural phenomena. This versatility is in part due to
the supplementary constitutive relations that are necessary to
accommodate the wide range of materials that respond to
electromagnetic fields. Although in principle such relations can
be derived from the underlying quantum description of matter, in
many practical situations one must rely on experimental guidance
to ascertain the classical response of materials to such fields.

Once the unifying power of a spacetime formulation of physical
phenomena became apparent with Einstein's relativistic world view,
the natural mathematical tool for describing  constitutive
responses became the total \SEM tensor  for all matter and fields.
Early suggestions by Minkowski  \cite{mink} and Abraham \cite{abr}
for the structure of its \EM component in simple media initiated a
long debate involving both theoretical and experimental
contributions that continues to the current time (see e.g.
\cite{peierls1}), \cite{peierls}, \cite{gordon}, \cite{brevik},
\cite{loudon},  \cite{ob_hehl}, \cite{feigel}, \cite{bowyer},
\cite{cambell}). Although it is widely recognised  that this
controversy is an argument about definitions \cite{feld} and that
the relative merits  of alternative definitions are undecidable
without a complete (experimentally verifiable) covariant
description of relativistic continuum mechanics for matter and
fields, it remains important to clarify the many conflicting
arguments that have appeared over the years and to offer new
insights that may help in modelling the electromagnetic properties
of moving media in the absence of a viable or complete description
of field-particle interactions at a more fundamental level.

Some way towards this goal is offered by (covariant) averaging
methods \cite{degrootbook}, \cite{degroot}. These however yield
non-symmetric \SEM tensors for \EM fields in simple media. If the
total \SEM is to remain symmetric this implies that other
asymmetric contributions must compensate and no guidance is
offered to account for such material induced asymmetries. The need
for a symmetric total \SEM tensor is often attributed to
conservation of total angular  momentum despite the fact that such
global conservation laws may not exist in arbitrary gravitational
fields. Although the magnitude of gravitational interactions may
be totally insignificant compared with the scale of those due to
electromagnetism, gravity does have relevance in establishing the
general framework (via the geometry of spacetime) for classical
field theory and in particular this framework \cite{israel},
\cite{mikura} offers the most cogent means to define the total
\SEM tensor as the source of relativistic gravitation. This in
turn may be related to a variational formulation
\cite{israelstewart} of the fully coupled field system of
equations that underpin the classical description of interacting
matter in terms of tensor  (and spinor) fields on spacetime.

In this article \SEM tensors are defined as variational
derivatives and it is argued that the classical properties of a
simple polarisable medium may be parameterised in terms of a
constitutive tensor whose properties can in principle be
determined by experiments in non-inertial (accelerating) frames
and in the presence of weak but variable gravitational fields.

There has been a rapid development in recent years in the
construction of \lq\lq traps" for confining collective states of
matter on scales intermediate between macro- and micro-dimensions.
Cold atoms and nano-structures offer many new avenues for
technological development when coupled to probes by \EM fields.
The constitutive properties of such novel material will play an
important role in this development. Space science is also
progressing rapidly and can provide  new laboratory environments
with variable gravitation and controlled acceleration in which the
properties of such states of matter may be explored. It will be
shown below that the response of electromagnetically polarisable
media to such novel experimental environments offers a means to
describe their \EM constitutive properties and hence gain insight
into the \EM stresses induced by \EM fields in such media.
Supplemented with additional data based on their mechanical and
elasto-dynamic responses one thereby gains a more confident
picture of the total phenomenological \SEM for media than that
based on previous ad-hoc choices.

Throughout this article the formulation will be expressed in terms
of tensor fields on spacetime with an arbitrary metric. Attention
will be drawn to conservation laws when this metric admits
particular symmetries. Thus the results have applicability to
simple media in arbitrary gravitational fields and accommodate
both media and observers with arbitrary velocities.


After establishing some geometric notation, section 2 relates the
electromagnetic 1-forms $\Me,\Mb,\Md,\Mh$ to the 2-forms $F$ and
$G$ that enter into Maxwell's phenomenological covariant field
equations in the presence of matter. Section 3 discusses stress,
energy and momentum {\it in the vacuum} and introduces the useful
notion of a {\it drive} form that can be used to calculate
electromagnetically induced  currents involving the flux of
electromagnetic energy, momentum and angular momentum in Minkowski
spacetime. In section 4 the definition of the \SEM tensor is
discussed with particular reference to its symmetry based on its
role as a source of relativistic gravitation. The constitutive
properties of the media considered in this paper are delineated in
section 5 in terms of a constitutive tensor on spacetime. This
includes an account of general {\it magneto-electric continua} and
leads in section 6 to a formulation of a self-adjoint constitutive
tensor describing, in general, inhomogeneous, anisotropic,
magneto-electric matter in arbitrary motion. The question of an
invariant characterisation of magnto-electric media is mentioned
in section 7. In section 8 an action principle is established to
generate the phenomenological Maxwell system and the use of
variational derivatives  to calculate \SEM tensors is discussed in
section 9. The computation of the electromagnetic \SEM tensor,
based on the action of section 8, is non-trivial for general media
exhibiting anisotropy and magneto-electric properties in arbitrary
motion and is presented in some detail. The relation of this
result to tensors proposed by Abraham and others is discussed in
the concluding section where the relevance of the whole approach
to experiments on matter in non-inertial environments with
variable gravitational and electromagnetic fields is stressed.


Notations follow standard conventions with spacetime modelled as a
4-dimensional,  orientable,  manifold $\man$ with a metric tensor
field $g$ of Lorentzian signature $(-,+,+,+)$. $\GamTM$ denotes
the set of vector fields and $\GamLamM{p}$ the set of $p-$form
fields on $\man$. The set $\Set{e^0,e^1,e^2,e^3}$ denotes a  local
$g$-orthonormal  coframe (a linearly independent collection of
$1-$ forms) with dual frame $\Set{X_0,X_1,X_2,X_3}$. If
$g_{ab}=g(X_a,X_b)$, the interior contraction operator $i_{X_a}$
with respect to $X_a$ is written $i_a$ with $i^a=g^{ab}i_{X_b}$,
$e_b= g_{ac} e^c$ and summation over $0,1,2,3$. Metric duals with
respect to $g$ are written with a tilde so that
$\dual{X}=g(X,-)\in\GamLamM{1}$ for $X\in\GamTM$ and
$\dual{\alpha}=g^{-1}(\alpha,-)\in\GamTM$ for
$\alpha\in\GamLamM{1}$. The Hodge dual map associated with $g$ is
denoted $\star$. The following standard identities will be used
repeatedly in subsequent sections to simplify expressions.
\begin{align}
\Phi\wedge\Psi = (-1)^{pq} \Psi\wedge\Phi \qquad&\text{for}\quad
\Phi\in\GamLamM{p},\ \Psi\in\GamLamM{q} \label{id_wedge}\\
\Phi\wedge\star\Psi=\Psi\wedge\star\Phi \qquad&\text{for}\quad
\Phi,\Psi\in\GamLamM{p} \label{id_star_pivot}
\\
i_X\star \Phi=\star(\Phi\wedge\tilde X)
\qquad&\text{for}\quad
X\in\GamTM,\
\Phi\in\GamLamM{p}
\label{id_iX_star}
\\
\star\, i_X\Phi=-\star\Phi\wedge\tilde X \qquad&\text{for}\quad
X\in\GamTM,\ \Phi\in\GamLamM{p} \label{id_star_iX}
\\
\star\star\Phi=(-1)^{p+1}\Phi
\qquad&\text{for}\quad
\Phi\in\GamLamM{p}
\label{id_star_star}
\\
i_X \Phi\wedge\Psi =
(-1)^{p+1} \Phi\wedge i_X\Psi
\quad&\text{for}\quad
\Phi\in\GamLamM{p},\
\Psi\in\GamLamM{q},\
p+q\ge5
\label{id_iX_move}
\\
d \Phi\wedge\Psi = (-1)^{p+1} \Phi\wedge d\Psi
\qquad&\text{for}\quad \Phi\in\GamLamM{p},\ \Psi\in\GamLamM{q},\
p+q\ge4 \label{id_d_move}
\end{align}

\def\ee{\,\epsilon_0\,}
\def\cc{\, c \,}

\section{Electromagnetic Fields}
\label{ch_fields}

Maxwell's equations for an electromagnetic field in an arbitrary
medium can be written
\begin{align}
d\,F=0 \qquadand d\,\star\, G =j
\label{intro_Maxwell}
\end{align}
where $F\in\GamLamM{2}$ is the Maxwell 2-form, $G\in\GamLamM{2}$
is the excitation 2-form and $j\in\GamLamM{3}$ is the 3-form
electric current source \footnote{All tensors in this article have
dimensions constructed from the SI dimensions $[M], [L], [T], [Q]$
where $[Q]$ has the unit of the Coulomb in the MKS system. We
adopt $[g]=[L^2], [G]=[j]=[Q],\,[F]=[Q]/\ee$ where the
permittivity of free space $\epsilon_0$ has the dimensions $ [
Q^2\,T^2 M^{-1}\,L^{-3}] $ and $c$ denotes the speed of light in
vacuo}. In general, the effects of gravitation and
electromagnetism on matter are encoded in this system in $\star G$
and $j$.  This dependence may be non-linear and non-local. To
close this system, ``\EM constitutive relations'' relating $G$ and
$j$ to $F$ are necessary. In the following the medium will be
considered as containing polarisable (both electrically and
magnetically) matter with $G$ restricted to a real point-wise
linear function of $F$, thereby ignoring losses and spatial and
temporal material dispersion in all frames. Continua endowed with
such properties will be termed ``simple'' here. The electric
4-current $j$ will be assumed to describe (free) electric charge
and plays no role in subsequent discussions.

The electric field $\Me\in\GamLamM{1}$ and magnetic induction
field $\Mb\in\GamLamM{1}$  associated with $F$ are defined with
respect to an arbitrary {\it unit} future-pointing timelike
$4-$velocity vector field $U\in\GamTM$ by
\begin{align}
\Me = \iuF \qquadand \cc\Mb = \iustarF \label{intro_e_b}
\end{align}
Since $g(U,U)=-1$
\begin{equation}
F=\Me\wedge \dualu - \star\,(\cc\Mb\wedge \dualu) \label{intro_F}
\end{equation}
The field $U$ may be used to describe an {\it observer frame} on
spacetime and its integral curves model idealised  observers.

Likewise the displacement field $\Md\in\GamLamM{1}$ and the
magnetic field $\Mh\in\GamLamM{1}$  associated with $G$ are
defined with respect to $U$ by
\begin{align}
\Md = \iu G\,, \qquadand \Mh/\cc = \iu\star G\,. \label{Media_d_h}
\end{align}
Thus
\begin{align}
G=\Md\wedge \dualu - \star\,((\Mh/\cc)\wedge \dualu)
\label{Media_G}
\end{align}
It will be assumed that a material medium has associated with it a
future-pointing timelike unit vector field $V$  which may be
identified with the  bulk $4-$velocity field of the medium in
spacetime. Integral curves of $V$ define the averaged world-lines
of identifiable constituents of the medium. A {\it comoving
observer frame with $4-$velocity $U$} will have $U=V$.


\section{Electromagnetic Stress, Energy and Momentum in the
Vacuum} \label{em_stress}

The historical development of Newtonian continuum mechanics led to
the notion of a stress tensor in Euclidean $3-$space that entered
into the balance laws for momentum and angular momentum. With the
advent of relativistic concepts this was generalised to a
stress-energy-momentum tensor in spacetime giving rise to
conserved quantities in situations where the metric admits
symmetries.

The basic properties of the electromagnetic stress-energy-momentum
tensor in the vacuum\footnote{The notion of a classical vacuum
here corresponds to spacetime devoid of all material ($j=0$)
although if $j$ has compact support one can refer to \lq\lq vacuum
domains" where $j=0$. All regions can admit non-zero \EM and
gravitational fields.}
 can be succinctly discussed in
terms of a set of  \lq\lq drive\rq\rq  $3-$forms. In vacuo the
Maxwell field system with a $3-$form electric current source $j$
satisfies
\begin{align}
d\,F=0 \qquadand \ee d\,\star\, F =j. \label{intro_Maxwell_vacuum}
\end{align}
For any vector field $\VV$ on spacetime and any Maxwell solution
$F$ to this system one can introduce a ``drive'' 3-form associated
with $Y$ and $F$
\begin{align}
\T{EM}_\VV=\frac{\epsilon_0}{2c}(i_\VV F \wedge \star\, F -
i_\VV\star\, F \wedge F)\label{ddrive}
\end{align}

This $3-$form can be used to generate different types of conserved
quantities when the vector field $\VV$ generates {\it (conformal)
isometries} on spacetime.\footnote{i.e. In terms of the Lie
derivative ${\cal L}_Y$, ${\cal L}_Y g=\lambda\,g$ for some scalar
$\lambda$. $Y$ is a Killing field when $\lambda=0$. Angular
momentum currents follow in terms Killing vector fields that
generate rotational diffeomorphisms. } If $\KK$ is any (conformal)
Killing vector on a domain of spacetime it then follows simply
from the vacuum Maxwell-system above that
\begin{align}
d\,\T{EM}_\KK=-\frac{1}{\cc}\,i_\KK\,F\wedge j
\end{align}
Thus for each (conformal) Killing vector field these equations
describe a ``local conservation equation'' ( $d\tau_\KK=0$) in a
source-free region $(j=0)$.

For $K$ any {\it unit timelike} Killing vector one has from
(\ref{ddrive})
\begin{align}
\T{EM}_K=\frac{1}{c^2} \Me \wedge  \Mh \wedge \tilde K -
\frac{1}{2c} \{\ee g(\tilde\Me,\tilde\Me) +\mu_0\,
g(\tilde\Mh,\tilde\Mh)\} \star\,\dual{K}
\end{align}
where $\Mh=\mu_0^{-1}\Mb$, $\Mb,\Me$ are defined with respect to
$U=K$ and $\mu_0\equiv \frac{1}{c^2\ee}$. The spatial $2-$form $
\Me\wedge \Mh$ was identified by Poynting in a source-free region
as proportional to the local field energy transmitted normally
across unit area per second (field energy current or power) and
$\frac{1}{2} \{\ee g(\tilde\Me,\tilde\Me) +
\mu_0\,g(\tilde\Mh,\tilde\Mh)\}$ proportional to the local field
energy density.
 More precisely $\int_\Sigma \tau_K$ is the field energy
associated with the spacelike 3-chain $\Sigma$ and
 $\int_{S^2}\,i_K\tau_K$ is the power flux across an oriented
spacelike 2-chain $S^2$.

If $X$ is a {\it unit spacelike} Killing vector generating
spacelike translations along open integral curves then with the
split:
\begin{align*}
\T{EM}_X=\mu_X \wedge \tilde V + \GX
\end{align*}
where $\iv\GX=0$, the Maxwell stress 2-form $\mu_X$ may be used to
identify mechanical forces produced by a flow of field momentum
current or pressure with momentum density 3-form $\GX$ \cite{liu}.

It is important to stress that different timelike Killing vectors
give rise to physically distinct notions of conserved energy. For
completeness the interpretation of \lq\lq energy"  requires
further information related to its mode of detection. The
existence of timelike {\it parallel} Killing vector fields
(including those whose integral curves are geodesics) are further
conditions that single out particular classes that may have
priority in establishing appropriate notions of conserved energy.

In general, in the absence of Killing vectors one loses strictly
conserved currents (closed $3-$forms) but a set of four local
$3-$form currents $\T{EM}_c\equiv \T{EM}_{X_c}$ can be defined in
any local coframe. In any frame $\{X_a\}$ with dual coframe
$\{e^b\}$ the 16 functions $T^{EM}_{ab}=i_{X_b} \star\,
\T{EM}_{a}$ may be used to construct the tensor
\begin{align}
T^{EM}=T^{EM}_{ab} \,e^a\otimes e^b
\end{align}
usually referred as the {\it stress-energy-momentum tensor}
associated with the above drive forms \footnote{ In view of the
above comments on the role of particular timelike and spacelike
Killing vectors in constructing conserved energy-power and
momentum-force currents a more coherent label for $T$ might be the
drive tensor} .

 The relationships between
any stress-energy tensor $T$ and the associated drive forms
$\tau_a$ are given by
\begin{align}
\tau_a = \star (T(X_a,-)) \qquadand T = \star(\tau_a\wedge e_b)
e^a\otimes e^b \label{intro_def_tau_a}
\end{align}
In terms of the $3-$forms $\tau_c$ the  symmetry condition  $
T_{bc}=T_{cb}$ is
\begin{equation}
e_c\wdd \tau_b=e_b \wdd \tau_c
\end{equation}

\section{The total \SEM tensor}\label{total_SEM}

When spacetime contains domains with matter (where $j$ may or may
not be zero) such regions will in general have physical properties
distinct from vacuum domains.

If a coupled system of electromagnetic, gravitational and matter
fields has a {\it total} stress-energy-momentum tensor
\begin{align}
T^{Total}=T^{EM}+ T^{m+I}
\end{align}
where $ T^{m+I} $ describes matter and its interactions not
included in $T^{EM}$, then on general grounds, if $T^{Total}$ is
symmetric,  one has:
\begin{equation}
{\nabla\cdot T^{Total}=0 }
\end{equation}
in terms of a (Koszcul) connection $\nabla$ on spacetime.
Different authors partition the total stress-energy tensor into a
sum of partial stress-energy tensors in different ways. The
divergences of certain partial stress-energy tensors are sometimes
called {\it pondermotive} forces.

If the connection $\nabla$,  induced from a connection on the
bundle of linear frames over spacetime,  is both metric compatible
and torsion free, with gravitational fields satisfying Einstein's
equations, then $T^{Total}$ must give rise to a symmetric
stress-energy tensor $T^{Total}_{ab}=T^{Total}_{ba}$. However any
such symmetric tensor can be partitioned into non-symmetric
partial tensors in infinitely many ways. Such a partition is then
an expediency without fundamental significance.

In theories of gravitation based on non-pseudo-Riemannian
geometries the {\it natural connection} may have torsion. For
example in an Einstein-Cartan theory with matter that gives rise
to a connection with torsion, the generalised Einstein tensor
$Ein^{EC}$, determined by varying the generalised Einstein-Hilbert
action with respect to orthonormal coframes, is non-symmetric and
hence the source tensor $T^{EC}$ defined by
\begin{equation}
{ Ein^{EC}=T^{EC}\label{EC}}
\end{equation}
is similarly non-symmetric. However for some forms of
gravitational-matter couplings the variation of the total action
with respect to the connection gives rise to algebraic equations
for the connection.\footnote{For example, locally $SL(2,C)$
covariant couplings of spinor fields to gravitation fall into this
category.} In principle these can be solved for the connection
which can always be decomposed into a sum containing the
torsion-free metric-compatible (Levi-Civita) connection used in
Einstein's pseudo-Riemannian description of gravitation. The
generalised Einstein tensor $Ein^{EC}$ can then be written $Ein+S$
in terms of the Einstein tensor $Ein$ and (\ref{EC}) becomes
\begin{equation}
{Ein=T^{E} \label{Ein}}
\end{equation}
where $T^{E}\equiv T^{EC}-S$ is symmetric and divergenceless with
respect to the Levi-Civita divergence. In such cases one may
define the total stress-energy tensor as the source tensor for
Einstein's equations (\ref{Ein}). It is then by definition
symmetric. If the natural connection $\nabla$ (determined by a
connection variation of the total action) gives rise to dynamic
torsion, determined by a partial differential system involving all
fields, the reduction to a geometrical formulation in terms of a
metric and Levi-Civita connection becomes an impracticality. In
such a situation the definition of the stress-energy tensor is
best left as $T^{EC}$. This has two distinct divergences with
respect to $\nabla$ since it is not symmetric.

Such general considerations offer guidance in the construction of
phenomenological partial stress-energy tensors based on either
coarse-graining detailed interactions between fields or the
introduction of effective degrees of freedom \cite{nelson}. Indeed
such phenomenological stress-energy tensors are often of greater
value than actions based on ``fundamental fields'' since they can
often be related more directly to experiment. Thus although in
this article gravitation will be regarded as a background
interaction
 the electromagnetic properties of a simple medium  will be accommodated into
certain constitutive tensors that respond to gravitation. We then
demand that an action describing such a medium in the absence of
free charges give rise by variation to Maxwell's phenomenological
equations for a simple medium {\it and} a symmetric stress-energy
tensor.

\section{The constitutive tensor for simple media}
\label{ch_Media}

In general $G$ may be a functional of $F$ and properties of the
medium\footnote{ e.g. electrostriction and magnetostriction arise
from the dependence of $ {\cal Z} $ on the elastic deformation
tensor of the medium \cite{antoci}.}.
\begin{align}
G = {\cal Z}[F,\ldots] \label{Media_Nonline_Constitutive}
\end{align}
Such a functional induces, in general, non-linear and non-local
relations between $\Md, \Mh$ and $\Me, \Mb$. These relations may
be explored either empirically or by coarse graining a suitable
macroscopic model. For general {\it linear continua}  one may have
for some positive integer $N$ and collection of {\it constitutive
tensor fields}
 $Z^{\,(r)}$ on spacetime the relation
\begin{align}
G = \Sigma_{r=0}^{N}{ Z^{\,{(r)}}}(\nabla^{\,r} F,\ldots)
\label{Media_General_Constitutive}
\end{align}
in terms of some spacetime connection $\nabla$. Additional
arguments refer to  variables independent of $F$ and its
derivatives. In this article, for the simple linear media under
consideration, we restrict to $$G=Z(F)$$ for some { constitutive
tensor field} $Z$ . In the vacuum $G=\epsilon_0 F$.

 A particularly simple linear isotropic medium may be described by a
bulk $4-$velocity field  $V$, a relative permittivity  scalar
field $\epsilon$ and a non-vanishing  relative permeability scalar
field  $\mu$. In this case $Z$ follows from
\begin{align}
\frac{G}{\ee} = &\epsilon \,\ivF\wedge\dualv -
\mu^{-1}\star(\ivstarF\wedge\dualv)\\= & (\epsilon -
\frac{1}{\mu})\, i_V F \wedge\dualv - \frac{1}{\mu} \, F
\label{Media_scalar_medium_G}
\end{align}
In a comoving frame with $U=V$  (\ref{Media_scalar_medium_G})
becomes
\begin{align}
\Md = \ee\epsilon\, \Me \qquadand \Mh = (\mu_0\mu)^{-1} \Mb \,
\label{Media_scalar_medium_dh}
\end{align}


For a non-magneto-electric but anisotropic medium, the relative
permittivity $\epsilon$ and inverse  relative permeability
$\mu^{-1}$ become {\it spatial tensor fields} on spacetime.  Thus
$\epsilon:\GamLamM{1}\to\GamLamM{1}$ and
$\mu^{-1}:\GamLamM{1}\to\GamLamM{1}$ for all $\alpha\in
\GamLamM{1}$ where
\begin{align}
\epsilon(\dualv)=0\,,\quad i_V\epsilon(\alpha)=0\,,\quad
\mu^{-1}(\dualv)=0\quadand i_V\mu^{-1}(\alpha)=0\,.
\label{Media_eps_mu_V}
\end{align}
The more general constitutive relation is then given by
\begin{align}
\frac{G}{\ee} = \epsilon(\ivF)\wedge\dualv -
\star(\mu^{-1}(\ivstarF)\wedge\dualv)
\label{Media_tensor_medium_G}
\end{align}
which in the comoving frame with  $U=V$
becomes
\begin{align}
\Md = \ee\epsilon(\Me) \qquadand \mu_0\Mh = \mu^{-1}(\Mb) \,.
\label{Media_tensor_medium_dh}
\end{align}
Based on standard thermodynamic arguments the inverse relative
permeability and relative permittivity tensors are symmetric with
respect to the metric $g$:
\begin{align}
i_{\dual{\alpha}}\epsilon(\beta)=i_{\dual{\beta}}\epsilon(\alpha)
\quadand
i_{\dual{\alpha}}\mu^{-1}(\beta)=i_{\dual{\beta}}\mu^{-1}(\alpha)
\quad\text{for}\quad  \alpha,\beta \in \GamLamM{1}
\end{align}
In general, the electromagnetic fields may be related by
\begin{equation}
\begin{aligned}
\Md =& \PermDE(\Me) + \PermDB(\Mb)
\\
\Mh =& \PermHE(\Me) + \PermHB(\Mb)
\end{aligned}
\label{Media_Constitutive_dh}
\end{equation}
where $\PermDE,\PermDB,\PermHE,\PermHB:\GamLamM{1}\to\GamLamM{1}$
are spatial tensors satisfying:
\begin{align}
&\PermGen(\dualv)=
\iv(\PermGen(\alpha))=0
\label{Media_iv_Perm_0}
\end{align}
and therefore
\begin{align}
\PermGen(\piV(\alpha))=
\piV(\PermGen(\alpha))=\PermGen(\alpha)
\label{Media_piV_Perm_0}
\end{align}
for $\PermGen=\PermDE,\PermDB,\PermHE,\PermHB$ and for
 all $\alpha\in \GamLamM{1}$,
where $\piV$ projects spacetime 1-forms to spatial 1-forms with
respect to $V$,  on spacetime
\begin{align}
\piV:\GamLamM{1}\to\GamLamM{1}\,,\qquad
\piV=\Id + \dualv\otimes V
\label{Media_def_pi_V}
\end{align}
From (\ref{Media_tensor_medium_dh}), (\ref{Media_Constitutive_dh})
it follows that if $\PermHE=\PermDB=0$ in some frame then
$\PermDE=\ee\epsilon$ and $\PermHB=(\mu_0\mu)^{-1}$ in that frame.
For such materials,however, one cannot assert that  $\PermHE,
\PermDB$ remain zero in all frames. Media with constitutive
relation (\ref{Media_Constitutive_dh}) are often referred to as
{\it magneto-electric} \cite{odell}. We prefer to use this term to
describe intrinsic magneto-electric media and will return to this
point in section \ref{ch_Post}.

The
tensor fields $\PermDE$, $\PermDB$, $\PermHE$ and $\PermHB$ are
encoded into the tensor $\PermTen:\GamLamM{2}\to\GamLamM{2}$ such
that $G=Z(F)$. Since
\begin{align}
\PermTen(\alpha+\beta)=\PermTen(\alpha)+\PermTen(\beta) \qquadand
\PermTen(\lambda \alpha)=\lambda\PermTen(\alpha)
\label{Media_Z_linear}
\end{align}
for all $\lambda\in\GamLamM{0}$ and $\alpha,\beta\in\GamLamM{2}$
the constitutive relation may be expanded in a local  co-frame
field  $\Set{e^0,e^1,e^2,e^3}$ as
\begin{align}
\tfrac12 G_{a b}e^a\wedge e^b = \tfrac14 \PermTen^{c d}{}_{a b}
F_{c d} e^a\wedge e^b
\end{align}
where
\begin{align}
\PermTen^{c d}{}_{a b} = - \PermTen^{c d}{}_{b a} = - \PermTen^{d
c}{}_{a b} = \PermTen^{d c}{}_{b a}
\label{Media_Z_abcd_Z_bacd}
\end{align}
These conditions alone imply that the tensor $\PermTen$ has 36
independent components, although additional symmetry conditions
given below will reduce these to 21. From the definition of $G$ in
terms of comoving fields and (\ref{Media_Constitutive_dh}), the
relationship between $\PermTen$ and
$\Set{\PermDE,\PermDB,\PermHE,\PermHB}$ follows as
\begin{equation}
\begin{aligned}
\PermTen(F) =& \PermDE(\ivF)\wedge\dualv +
\PermDB(\ivstarF)\wedge\dualv
\\&
- \star(\PermHE(\ivF)\wedge\dualv) - \star(\PermHB(\iv \star
F)\wedge\dualv)
\end{aligned}
\label{Media_Z_decomp}
\end{equation}
and hence by contraction with $V$
\begin{equation}
\begin{aligned}
& \PermDE(\xi) = \iv \PermTen(\xi\wedge\dualv) \,,\quad
\PermDB(\xi) = - \iv \PermTen(\star(\xi\wedge\dualv)) \,,\quad
\\&
\PermHE(\xi) = \iv\star \PermTen(\xi\wedge\dualv) \,,\quad
\PermHB(\xi) = - \iv\star \PermTen(\star(\xi\wedge\dualv))
\end{aligned}
\label{Media_def_Z_emme}
\end{equation}

\section{Symmetry of the constitutive tensor.}\label{Z_SYM}

The adjoint of any tensor  $T:\GamLamM{p}\to\GamLamM{p}$, is the
tensor $T^\dagger:\GamLamM{p}\to\GamLamM{p}$ defined by
\begin{align}
\alpha\wedge\star T(\beta) = \beta\wedge\star T^\dagger(\alpha)
\qquad\text{for }\alpha,\beta\in\GamLamM{p}
\label{Media_def_adjoint}
\end{align}
Clearly $T^{\dagger\dagger}=T$. If $p=1$,
(\ref{Media_def_adjoint}) gives
\begin{align}
i_{\dual{\alpha}} T(\beta) = i_{\dual{\beta}} T^\dagger(\alpha)
\qquad\text{for }\alpha,\beta\in\GamLamM{1}
\label{Media_def_adjoint_1forms}
\end{align}
The symmetry conditions for the relative permittivity and inverse
permeability tensors imply that $\PermDE$ and $\PermHB$ are self
adjoint. This symmetry is generalised to magneto-electric media:
\begin{align}
\PermDE^\dagger=\PermDE \,,\quad \PermHB^\dagger=\PermHB \quadand
\PermDB^\dagger=-\PermHE \label{Media_Zeta_adjoint}
\end{align}
i.e.  $\PermTen$ is assumed self-adjoint
\begin{align}
\PermTen=\PermTen^\dagger\, \label{Media_Z_adjoint}
\end{align}
or, raising indices with the metric
\begin{align}
Z^{abcd}=Z^{cdab}
\label{Media_Z_adjoint_abcd}
\end{align}
Using sequentially
(\ref{Media_def_Z_emme}), (\ref{id_star_iX}), (\ref{Media_Z_adjoint}),
(\ref{id_iX_star}), (\ref{id_iX_move}), (\ref{Media_def_Z_emme}),
(\ref{id_wedge}), (\ref{id_star_pivot}), (\ref{Media_def_adjoint})
this condition yields
\begin{align*}
\alpha\wedge\star\PermDB(\beta) =& -\alpha\wedge\star\iv
\PermTen(\star(\beta\wedge\dualv)) = \alpha\wedge\dualv\wedge\star
\PermTen(\star(\beta\wedge\dualv))
\\
=& {\star}(\beta\wedge\dualv)\wedge\star
\PermTen(\alpha\wedge\dualv) =
\iv{\star}\beta\wedge{\star}\PermTen(\alpha\wedge\dualv) =
\star\beta\wedge\iv{\star}\PermTen(\alpha\wedge\dualv)
\\
=& \star\beta\wedge\PermHE(\alpha) =
-\PermHE(\alpha)\wedge\star\beta =
-\beta\wedge\star\PermHE(\alpha)= -\alpha\wedge\star
\PermHE^{\dagger}(\beta)
\end{align*}
i.e. $\PermDB=-\PermHE^\dagger$.  The remaining equations in
(\ref{Media_Zeta_adjoint}) follow similarly.

It follows from (\ref{Media_Z_abcd_Z_bacd}) and
(\ref{Media_Z_adjoint_abcd}) that the number of independent components of
$Z$ reduce from 36 to 21.

\section{Intrinsic magneto-electric media and the Post constraint.}
\label{ch_Post}

A constitutive tensor $\PermTen$ describes a {\it non
  intrinsic-magneto-electric medium} if there exists a velocity field
$V$ for the medium such that $\PermDB=0$ and $\PermHE=0$.  Thus a
constitutive tensor $\PermTen$  is intrinsically magneto-electric
if there does not exists a velocity field $V$ such that
$\PermDB=0$ and $\PermHE=0$.  If $Z(F)$ is decomposed with respect
to an arbitrary frame $U\ne V$ one may find all tensors $\zeta$
non-zero, even for media that are not intrinsically
magneto-electric. For a general constitutive tensor, it is a
matter of linear algebra to decide whether it  describes  an
intrinsically  magneto-electric  medium or not.

A useful characterisation of magneto-electric media may be given
in terms of invariants constructed from $Z$ and the metric. One
such invariant introduced by Post \cite{post},
\cite{ob_hehl_post}, is
\begin{align}
\chi = \ia\ib \star (\PermTen(e^a\wedge e^b))
\label{Media_Post_def}
\end{align}
In terms of spatial tensors with respect the the medium velocity
$V$
\begin{equation}
\begin{aligned}
\chi
=&
\ia\ib \star (\PermTen(e^a\wedge e^b))
\\
=&
\ia\ib{\star}(\PermDE(\iv (e^a\wedge e^b))\wedge\dualv)
+ \ia\ib \star(\PermDB(\iv{\star}(e^a\wedge e^b))\wedge\dualv)
\\&
+ \ia\ib (\PermHE(\iv(e^a\wedge e^b))\wedge\dualv)
+ \ia\ib (\PermHB(\iv{\star}(e^a\wedge e^b))\wedge\dualv)
\\
=& \ia\ib \star(\PermDB(\iv{\star}(e^a\wedge e^b))\wedge\dualv)
+ \ia\ib (\PermHE(\iv(e^a\wedge e^b))\wedge\dualv)
\end{aligned}
\label{Media_expand_chi}
\end{equation}
since
\begin{align*}
\ia\ib{\star}(\PermDE(\iv (e^a\wedge e^b))\wedge\dualv) =
\ia\ib (\PermHB(\iv{\star}(e^a\wedge e^b))\wedge\dualv) = 0
\end{align*}
Using
\begin{align*}
\star(e^a\wedge e^b\wedge \dualv)
= \star(e^a\wedge e^b\wedge e^c)V_c
= V_c \star(e^a\wedge e^b\wedge e^c\wedge e^d) e_d
= - \varepsilon^{abcd} V_c e_d
\end{align*}
and
\begin{align*}
\star(\xi\wedge\dualv\wedge e_b\wedge e_a)
= \star(e_a \wedge e_b \wedge \dualv \wedge\xi)
= \star(e_a \wedge e_b \wedge e_e \wedge e_f) V^e \xi^f
= -\varepsilon_{abef} V^e\xi^f
\end{align*}
with $\varepsilon_{abef} \varepsilon^{abcd} =
\delta^d_e\delta^c_f-\delta^c_e\delta^d_f$, the first term on the last
line of (\ref{Media_expand_chi}) yields
\begin{align*}
\ia\is \star(\PermDB(\iv{\star}(e^a\wedge e^s))\wedge\dualv) =&
\star(\PermDB(\star(e^a\wedge e^s\wedge \dualv))
\wedge\dualv\wedge e_s\wedge e_a)
\\
=& -\star(\PermDB(e_r)\wedge\dualv\wedge e_s\wedge
e_a)\varepsilon^{ascr} V_c
\\
=& i^f\PermDB(e_r) \varepsilon_{asef} \varepsilon^{ascr}V^e V_c =
2 \ia\PermDB(e^a)
\end{align*}
while the second term is
\begin{align*}
\ia\ib (\PermHE(\iv(e^a\wedge e^b))\wedge\dualv)
&=
\ia\ib(\PermHE(v^a e^b)\wedge\dualv) -\ia\ib(\PermHE(v^be^a)\wedge\dualv)
\\
&=
2\iv\ib (\PermHE(e^b)\wedge\dualv)
=
-2\ib\PermHE(e^b)
\end{align*}
Hence using (\ref{Media_Zeta_adjoint})
\begin{align}
\chi = 4\ia \PermDB(e^a) = -4\ia\PermHE(e^a)
\label{Media_Post_Zdb}
\end{align}
Thus since $V$ is  an arbitrary medium velocity,  a sufficient
condition for a medium to be intrinsically  magneto-electric is
that $\chi\ne 0$.

However some intrinsically magneto-electric media may have
$\chi=0$.  For example, consider the self-adjoint constitutive tensor given, in
some local orthonormal coframe $\{e^0, e^1,e^2,e^3\}$, by
\begin{align*}
\PermTen(F) =
  F_{23} e^0\wedge e^1
+ F_{13} e^0\wedge e^2 - F_{02} e^1\wedge e^3 - F_{01}
e^2\wedge e^3
\end{align*}
Then with $V=X_0$
\begin{align*}
\PermDB(\xi) = (i_1 \xi) e^1 - (i_2 \xi) e^2
\qquadand
\PermHE(\xi) = -(i_1 \xi) e^1 + (i_2 \xi) e^2
\end{align*}
and $\chi=0$. However one easily verifies that $\PermDB\ne0$ with
respect to any arbitrary unit timelike $V$. Hence $Z$ describes an
intrinsically magneto-electric medium.

A minimal set of invariants whose non-vanishing is a necessary and
sufficient condition for a medium to be intrinsically
magneto-electric is not known to the authors.


\section{Action for Source Free Electromagnetic Fields in a Simple Medium}
\label{ch_Action_Max}

The classical equations describing the total system of matter and
fields will be considered as arising from the extremum of some {\it
total action functional} under suitable variations with compact
support. This action should be constructed from an action density
4-form on spacetime in terms of (pull-backs) of sections (and their
derivatives) of field bundles carrying representations of local
symmetry (gauge) groups and maps between them. Observed local
symmetries in nature arise in such a formalism by ensuring that the
action $4-$form is a scalar under local changes of section. To
maintain these covariances appropriate connections are required to
define tensorial (and spinorial)  covariant derivatives of sections.
In addition the action may depend on tensor-valued functions of
these sections. All variational principles require a specification
of what objects in the action are to be varied and these then
constitute the dynamical variables of the theory. In the following
we concentrate on a contribution $\Lambda$ to the total action
arising from the effects of the electromagnetic field and
gravitation  in different types of \lq\lq media". We exclude from
this $\Lambda$ the interaction with charged matter and the dynamics
of the gravitational field itself. Included is the effect of the
electromagnetic field on a polarisable and magnetisable medium
assumed to be described in terms of a particular constitutive tensor
$Z$. In  particular we explore how the response of the medium to
gravitation as well as the electromagnetic field can be used to
establish the stress-energy-momentum tensor associated with
different choices of constitutive tensor. Thus the action 4-form
$\Lambda$  will be taken to depend only on the spacetime metric and
the class of Maxwell 1-form potentials $A$ with $F=dA$.  The
dependence of the tensor field $Z$ on these variables will be
explored in some detail below.

We have insisted  that in the absence of free charge the
electromagnetic fields $F$ and $G$ for a simple medium in any
spacetime metric must satisfy
\begin{equation}
d\,F=0 \qquadand d\star\, G=0
\label{Action_Max_Maxwells_equations}
\end{equation}
Before generating an electromagnetic \SEM  tensor from a
particular contribution to the total action it is necessary to
verify that these field equations arise by suitable variation.
Consider then the contribution  $S[A,g]=\int_M \Lambda$ where
$F=d\,A$ ,$G=\PermTen(F)$ with $\PermTen=\PermTen^\dagger$ and
\begin{equation}
\cc\Lambda = \tfrac{1}{2}\,F \wedge \star\, G = \tfrac12\,F \wedge
\star\, \PermTen(F) \, \label{Action_Max_action}
\end{equation}
If a  prime denotes the variation with respect to $A$, then
working modulo $d$:
\begin{align*}
\cc\Lambda' =& \tfrac12 \Big( d A'\wedge\star\PermTen( d A) + d
A\wedge\star\PermTen( d A') \Big)
\\
=& d A'\wedge\star\PermTen( d A) =
 A'\wedge d \star \PermTen( d A)
= A'\wedge d\star G
\end{align*}
Hence the source-free Maxwell equations
(\ref{Action_Max_Maxwells_equations}) follow by variation with
respect to $A$ from the action (\ref{Action_Max_action}). Note
that the symmetry condition (\ref{Media_Z_adjoint}) of the tensor
$\PermTen$ is essential in this variation.

\section{Variational derivatives and   \SEM tensors}
\label{ch_Metric}

To effect the metric variations  of the above action functional
let $t\to\gT$ be a curve in the space of Lorentzian signatured
metrics, with $\gO\equiv \gT|_{t=0}$. The ``tangent'' to the the
curve $t\to\gT$ at the point $t=0$ is written $\gD$:
\begin{align}
\gT = \gO + t \gD + O(t^2) \label{Metric_g_t_0_dot}
\end{align}
For a general object $K$ which may be a tensor or a map which
depends on the metric $g$, write similarly $t\to K_t$ as the one
parameter set of objects encoding the dependence of $K$ on $\gT$,
$K_0=K_t|_{t=0}$ and $\dot{K}=\tfrac{d}{dt}K_t|_{t=0}$, so
\begin{align}
K_t=K_0 + t \dot{K} + O(t^2) \label{Metric_general_lift}
\end{align}
 $K_t$ will be referred to as the metric induced  {\it{lift}} of $K$.

One may represent the local variation $\gT$ in different ways. One
way is to vary the components of $\gT$ with respect to a fixed
local co-frame $\{e^a_0\}$, i.e.
\begin{align}
\gT=(\gT)_{ab} e_0^a\otimes e_0^b \qquad\text{where}\quad
(\gT)_{ab} = \gT\big((X_0)_a,(X_0)_b\big) \label{Metric_gt_ab}
\end{align}
One can set the fixed frame to be orthonormal with respect to the
unvaried metric so that
$(\gO)_{ab}=\eta_{ab}=\text{diag}(-1,+1,+1,+1)$. The derivative
$\gD$ is therefore given by
\begin{align}
\gD=\gD_{ab} e_0^a\otimes e_0^b \label{Metric_gD_ab}
\end{align}
Alternatively  one may vary the co-frame: i.e. choose a one
parameter set of coframes $t\to e_t^a$ for $a=0,..,3$ such that
$e_t^a|_{t=0}=e_0^a$ and
\begin{align}
\gT=\eta_{ab} e_t^a \otimes e_t^b \label{Metric_et_a}
\end{align}
The derivative of  $t\to e_t^a$ at $t=0$ follows from
\begin{align}
e^a_t = e^a_0 + t\,\deltae^a + O(t^2) \label{Metric_eD_a}
\end{align}
The derivative $\gD$ may also therefore be written
\begin{equation}
\gD= \eta_{ab}\left( \dot e^a \otimes e^b_0 + e^a_0 \otimes \dot
e^b\right) \label{Metric_gD_eD}
\end{equation}
The drive 3-forms $\tau_a$  associated with any action 4-form
$\Lambda$ are defined by the variation of $\Lambda$  with respect
to the orthonormal coframe as
\begin{align}
\dot{\Lambda} = \deltae^a \wedge \tau_a \label{Lift_def_tau_a}
\end{align}
If  the variation of $\Lambda$ with respect to the ortho-normal
coframe is induced entirely from the metric $g$ (and the metric
compatible torsion-free Levi-Civita connection) then
\begin{align}
\tau_a = 2 i_{X_b} \left(\frac{\delta\Lambda}{\delta
g_{bc}}\right) \eta_{ac} \label{Metric_g_var_tau_a}
\end{align}


This follows immediately  by equating (\ref{Metric_gD_ab}) and
(\ref{Metric_gD_eD}):
\begin{align*}
\gD_{ab} e_0^a\otimes e_0^b= \eta_{ab}\left( \dot e^a \otimes
e^b_0 + e^a_0 \otimes \dot e^b\right)
\end{align*}
so
\begin{align*}
\gD_{ab} = \eta_{cd}\left( \dot e^c(X_a)\delta^d_b + \delta^c_a
\dot e^d(X_b)\right) = \dot e_a(X_b) + \dot e_b(X_a) =i_b\dot e_a
+ i_a\dot e_b
\end{align*}
since one may drop the $0$ subscript  here without ambiguity:
$X_a=(X_0)_a$. Then
\begin{align*}
\dot\Lambda =& \deltae^a \wedge \tau_a =
\frac{\delta\Lambda}{\delta g_{ab}} \gD_{ab} =
\frac{\delta\Lambda}{\delta g_{ab}} (\dot e_a(X_b) + \dot
e_b(X_a)) = 2 \frac{\delta\Lambda}{\delta g_{ab}} \dot e_a(X_b)
\\
=& 2 \dot e_a \wedge i_b \left(\frac{\delta\Lambda}{\delta
g_{ab}}\right)
\end{align*}
By (\ref{intro_def_tau_a}) the tensor associated  to the $\tau_a$
is given by
\begin{align}
T =-2\star \left(\frac{\delta\Lambda}{\delta g_{ab}}\right)
e_a\otimes e_b \label{Metric_g_var_tau_a_eb}
\end{align}
and is manifestly  symmetric.

In the following it is necessary to make explicit the metric
dependence of various elements that enter in the action $4-$form
$\Lambda$ and in particular to pass between vector fields and
forms using the varied metric $g_t$.
 Thus  the notations $\gT:\GamTM\to\GamLamM{1}$,
$X\mapsto\gT(X)$ and $\gT^{-1}:\GamLamM{1}\to\GamTM$,
$\alpha\mapsto\gT^{-1}(\alpha)$ for the metric dual of vectors and
1-forms with respect to the metric $\gT$ are used.
 For vectors or 1-forms which already have a
subscript $0$ or $t$ we  continue to use the tilde notation
without ambiguity so that, for example, the $\gT$ metric dual of
the vector $X_t$ can be written $\dual{X}_t\equiv\gT(X_t)$.

Following (\ref{Metric_general_lift}) one has the maps  $\starO$,
$\starT$ and $\starD$ and from the  Leibnitz rule (evaluated at $t=0$)
\begin{align}
(\star\alpha)\dot{}=\starD\alpha+\star\dot\alpha
\label{Metric_star_alpha_dot}
\end{align}
for all $\alpha_t\in\GamLamM{p}$. It follows simply (see appendix)
that
\begin{align}
\starD\alpha = \deltae^a\wedge\ia\star\alpha -
\star(\deltae^a\wedge\ia\alpha) \qquad\text{for}\quad
\alpha_t\in\GamLamM{p} \label{Lift_starD}
\end{align}
Taking the derivative of
$\Phi\wedge\starT\Psi=\Psi\wedge\starT\Phi$ with respect to $t$
gives
\begin{align}
\Phi\wedge\starD\Psi=\Psi\wedge\starD\Phi
\qquad&\text{for}\quad
\Phi,\Psi\in\GamLamM{p}
\label{id_starT_pivot}
\end{align}

Thus with the metric induced lift of the constitutive tensor
$\PermTen$:
\begin{align}
\PermTenT = \PermTenO + t\,\PermTenD + O(t^2)
\label{Lift_Z_t_0_dot}
\end{align}
one writes:
\begin{align}
(\star Z_t(F))\dot{} = \dot\star Z(F) + \star\dot Z(F)
\label{RWT1}
\end{align}

Since there is a one parameter set of Hodge duals, we need to
distinguish  $\daggerT$ and $\daggerO$. Furthermore
(\ref{Media_def_adjoint}) becomes
\begin{align}
\alpha\wedge\starT T(\beta) = \beta\wedge\starT T^\daggerT(\alpha)
\label{Media_def_adjoint_t0}
\end{align}
for all $\alpha,\beta\in\GamLamM{p}$ and
(\ref{Media_def_adjoint_1forms}) becomes
\begin{align}
i_{\gT^{-1}\alpha} T(\beta) = i_{\gT^{-1}\beta} T^\daggerT(\alpha)
\label{Media_def_adjoint_t0_1forms}
\end{align}
for all $\alpha,\beta\in\GamLamM{1}$.

\section{Computation of the \SEM tensor   } \label{ch_Mink}

In this section the variation of the above action (\ref
{Action_Max_action} )is explored for a particular  choice of  the
metric dependence for $Z_t$, corresponding to a  perturbative
response of the medium to gravitation.

For a general lift the action  $4-$form (\ref{Action_Max_action})
is written
\begin{align}
\cc\Lambda_t = \tfrac12 F \wedge \starT  Z_t (F)
\label{Mink_Lagrange}
\end{align}
hence
\begin{align}
\cc\dot\Lambda = \tfrac12\big( F\wedge\starD \PermTen(F) + F
\wedge \star  \dot{Z} (F) \big)
\end{align}

From (\ref{Lift_starD})
\begin{equation}
\begin{aligned}
F\wedge\starD \PermTen(F) =& F\wedge\starD G = F \wedge \deltae^a
\wedge \ia \star G - F \wedge \star (\deltae^a \wedge \ia G)
\\
=& \deltae^a \wedge F \wedge \ia \star G  - \deltae^a \wedge \ia G
\wedge \star F
\\
=& \deltae^a \wedge \big( F \wedge \ia \star G - \ia G \wedge
\star F \big)
\end{aligned}
\label{Mink_tau_a_Mink}
\end{equation}
and
\begin{align*}
F \wedge \star  \dot{Z} (F) = 2F\wedge \star \frac{\delta
Z}{\delta g_{ab}}(F) i_b\deltae^a = 2\deltae^a\wedge
i_b\left(F\wedge \star \frac{\delta Z}{\delta
  g_{ab}}(F)\right)
\end{align*}

Therefore the drive forms are given by:
\begin{align}
\cc\tau_a = \tfrac12F \wedge \ia \star G - \tfrac12\ia G \wedge
\star F + i_b \left( F\wedge\star\frac{\delta Z}{\delta
g_{ab}}(F)\right) \label{Mink_general_tau_a}
\end{align}

For a physical medium with bulk motion that can sustain elastic
stresses associated with its atomic constituents one expects that
the history of such bulk motion should have some influence on the
constitutive properties via some associated 4-velocity
field\footnote{Relativistic strings and membranes with dynamics
that arise from re-parameterisation independent actions are an
exception since,  without \lq\lq constituents",  no preferred
parametrisation of their histories should  be identified.}.
 To include the possible dependence of the stress-energy-momentum
tensor on  such bulk motion of the medium one  requires  $Z$ to
depend on this motion in some manner. In
(\ref{Media_Constitutive_dh}) $Z$ is specified in terms of
electromagnetic fields measured in the comoving frame $V$ of the
medium. It is therefore natural to prescribe a lift of this
expression involving the lifts of $V_0$ and
$\Set{\PermDEO,\PermDBO,\PermHEO,\PermHBO}$. The natural  lift of
the medium velocity $\vO$ is
\begin{align}
\vT = \frac{\vO}{\sqrt{-\gT(\vO,\vO)}} \label{Lift_V_t}
\end{align}
The metric dual of $\vT$ is given by $\dualvT = \gT(\vT)$ and the
projection $\piV$ (\ref{Media_def_pi_V})
is lifted to
\begin{align}
\piT=\Id + \dualvT\otimes\vT
\label{Lift_pi_t}
\end{align}

The decomposition (\ref{Media_def_Z_emme}) of $\PermTenO$ and
$\PermTenT$ with respect to the medium velocities $\vO$ and $\vT$
is given by $\Set{\PermDEO,\PermDBO,\PermHEO,\PermHBO}$ and
$\Set{\PermDET,\PermDBT,\PermHET,\PermHBT}$ respectively,
following the notation (\ref{Metric_general_lift}).


The lifted tensors $\PermDET$, $\PermDBT$, $\PermHET$ and
$\PermHBT$ will be now chosen to  satisfy three properties:

\begin{jglist}
\item For all $t$ in the neighbourhood of  $t=0$
\begin{align}
\PermGenT|_{t=0} = \PermGenO \qquad\text{for}\quad
\PermGenT=\PermDET,\PermDBT,\PermHET,\PermHBT
\label{Lift_Perm_t_0}
\end{align}
\item For all $t$ in the neighbourhood of  $t=0$ they map the
vector space that is $g_t-$orthogonal to $\dualvT$ to itself. This
is achieved by lifting (\ref{Media_iv_Perm_0}):
\begin{align}
&\PermGenT(\dualvT)=0 \quadand i_{\vT}\PermGenT(\alpha)=0
\label{Lift_iv_Perm_0}
\end{align}
for $\PermGenT=\PermDET,\PermDBT,\PermHET,\PermHBT$ and all
$\alpha\in\GamLamM{1}$. The corresponding lift of
(\ref{Media_piV_Perm_0}) is
\begin{align}
\PermGenT(\piT(\alpha))=
\piT(\PermGenT(\alpha))=\PermGenT(\alpha)
\label{Lift_piV_Perm_0}
\end{align}

\item For all $t$ in the neighbourhood of  $t=0$ they retain the
adjoint conditions (\ref{Media_Zeta_adjoint}).
\begin{align}
(\PermDET)^{\daggerT}=\PermDET \,,\quad
(\PermHBT)^{\daggerT}=\PermHBT \quadand
(\PermDBT)^{\daggerT}=-\PermHET \label{Lift_Perm_adjoint}
\end{align}
\end{jglist}
These requirements are all satisfied by setting
\begin{align}
i_X\PermGenT(\alpha) =\tfrac12 \left( i_X\PermGenO(\piT\alpha) +
i_{\gT^{-1}\alpha}(\PermGenO)^{\daggerO}(\piT\gt X) \right)
\label{Lift_PermGen_t_res2}
\end{align}
for all $\alpha\in\GamLamM{1}$ and $X\in\GamTM$, i.e.\footnote{
For an isotropic, non-magneto-electric medium
(\ref{Media_scalar_medium_G}) and (\ref{Media_scalar_medium_dh}),
the lifts (\ref{Lift_PermGen_t_res3}) reduces to the lifts
\begin{align*}
\PermDET=\epsilon\piT \,,\quad \PermHBT=\mu^{-1}\piT \,,\quad
\PermDBT=0 \quadand \PermHET=0
\end{align*}
which in a comoving frame yield the relations
\begin{align*}
\MdT = \ee\epsilon\, \MeT \qquadand \MhT = (\mu_0\,\mu)^{-1} \MbT
\,
\end{align*}
where the scalars  $\epsilon$ and $\mu^{-1}$ are independent of
the ambient metric.}

\begin{equation}
\begin{aligned}
i_X\PermDET(\alpha) =& \tfrac12\left( i_X\PermDEO(\piT\alpha) +
i_{\gT^{-1}\alpha}\PermDEO(\piT\gT X) \right)\,,
\\
i_X\PermHBT(\alpha) =& \tfrac12\left( i_X\PermHBO(\piT\alpha) +
i_{\gT^{-1}\alpha}\PermHBO(\piT\gT X) \right)\,,
\\
i_X\PermDBT(\alpha) =& \tfrac12\left( i_X\PermDBO(\piT\alpha) -
i_{\gT^{-1}\alpha}\PermHEO(\piT\gT X) \right)\,,
\\
i_X\PermHET(\alpha) =& \tfrac12\left( i_X\PermHEO(\piT\alpha) -
i_{\gT^{-1}\alpha}\PermDBO(\piT\gT X) \right)
\end{aligned}
\label{Lift_PermGen_t_res3}
\end{equation}

To verify that (\ref{Lift_PermGen_t_res2}) obeys (\ref{Lift_Perm_t_0})
note that from (\ref{Media_def_pi_V})
\begin{align*}
\gO^{-1}\piO\gO X = X + \gO(X,\vO)\vO
\end{align*}
and hence from (\ref{Media_def_adjoint_t0_1forms}) and
(\ref{Media_iv_Perm_0})
\begin{align*}
i_{\gO^{-1}\alpha}(\PermGenO)^{\daggerO}(\piO\gO X)
= i_{\gO^{-1}\piO\gO X}\PermGenO(\alpha) = i_X\PermGenO(\alpha)
\end{align*}
Thus at $t=0$ (\ref{Lift_PermGen_t_res2}) becomes
\begin{align*}
i_X\PermGenT(\alpha)|_{t=0} =& \tfrac12 \left(
i_X\PermGenO(\piO\alpha) +
i_{\gO^{-1}\alpha}(\PermGenO)^{\daggerO}(\piO\gO X) \right)
=i_X\PermGenO(\alpha)
\end{align*}
using (\ref{Media_piV_Perm_0}).

To verify that (\ref{Lift_PermGen_t_res2}) obeys
(\ref{Lift_iv_Perm_0})
observe that
\begin{align*}
i_X\PermGenT(\dualvT) =\tfrac12 \left( i_X\PermGenO(\piT\dualvT) +
i_{\gT^{-1}\dualvT}(\PermGenO)^{\daggerO}(\piT\gt X) \right)
\end{align*}
Now $\piT\dualvT=0$ so the first term vanishes. Also using
(\ref{Media_def_adjoint_t0_1forms})
\begin{align*}
i_{\gT^{-1}\dualvT}(\PermGenO)^{\daggerO}(\piT\gt X)
=&
\ivT (\PermGenO)^{\daggerO}(\piT\gt X)
=
\sqrt{-\gT(\vO,\vO)}\
i_{\vO}(\PermGenO)^{\daggerO}(\piT\gt X)
\\
=&
\sqrt{-\gT(\vO,\vO)}\  i_{\gT^{-1}\piT\gT X}\PermGenO(\dualvO)
=0
\end{align*}
Hence $\PermGenT(\dualvT)=0$. Likewise
\begin{align*}
\ivT \PermGenT(\alpha) = \tfrac12 \left( \ivT
\PermGenO(\piT\alpha) +
i_{\gT^{-1}\alpha}(\PermGenO)^{\daggerO}(\piT\dualvT) \right) = 0
\end{align*}

Finally to verify that (\ref{Lift_PermGen_t_res3}) obeys
(\ref{Lift_Perm_adjoint}) use
(\ref{Media_def_adjoint_t0_1forms}) and (\ref{Lift_PermGen_t_res3}) twice
\begin{align*}
i_{\gT^{-1}\alpha}(\PermDBT)^{\daggerT}(\beta) =&
i_{\gT^{-1}\beta}\PermDBT(\alpha) = \tfrac12\left(
i_{\gT^{-1}\beta}\PermDBO(\piT\alpha) -
i_{\gT^{-1}\alpha}\PermHEO(\piT\beta) \right)
\\
=& - i_{\gT^{-1}\alpha}\PermHET(\beta)
\end{align*}
In a similar way it follows that $(\PermDET)^{\daggerT}=\PermDET$
and $(\PermHBT)^{\daggerT}=\PermHBT$. Thus
(\ref{Lift_PermGen_t_res3}) provide natural conditions for the
lifts (\ref{Lift_Perm_t_0}) to (\ref{Lift_Perm_adjoint})\footnote{
The requirements (\ref{Lift_Perm_t_0}-\ref{Lift_Perm_adjoint})
are not meant to be exhaustive. Other lifts $\PermGenT$ could
involve gradients of the spacetime metric corresponding to
gravitational tidal effects on the constitutive tensor. For
example if $\cal R$ is  the curvature scalar associated with the
Levi-Civita connection  then the lifts
\begin{align*}
i_X\PermGenT(\alpha) =\tfrac12({\cal R}_t - {\cal R}_0 + 1) \left(
i_X\PermGenO(\piT\alpha) +
i_{\gT^{-1}\alpha}(\PermGenO)^{\daggerO}(\piT\gt X) \right)
\end{align*}
also satisfy (\ref{Lift_Perm_t_0}-\ref{Lift_Perm_adjoint}).}.


Inserting the relations (\ref{Lift_PermGen_t_res3}) into $Z(F)$,
(\ref{Media_Z_decomp}), the action $4-$form  (\ref{Mink_Lagrange})
becomes
\begin{equation}
\begin{aligned}
2\cc\LagT =& F\wedge\starT(\PermDET(\ivT F)\wedge\dualvT) +
F\wedge\starT(\PermDBT(\ivT\starT F)\wedge\dualvT)
\\
&+ F\wedge\PermHET(\ivT F)\wedge\dualvT + F\wedge\PermHBT(\ivT
\starT F)\wedge\dualvT \,
\end{aligned}
\label{Lift_Lag_T}
\end{equation}
To ease the density of notation  in the following, the symbol
$\MbT$ now stands for $\cc\MbT$  and $\MhT$ stands for
$\frac{\MhT}{c}$. The lifts
\begin{align}
\MeT=\ivT F \,,\quad \MbT=\ivT\starT F \,,\quad
\dualMeT=\gT^{-1}(\ivT F) \quadand \dualMbT=\gT^{-1}(\ivT\starT F)
\label{lift_eT_bT}
\end{align}
satisfy
\begin{align}
\piT\MeT=\MeT
\qquadand
\piT\MbT=\MbT
\label{lift_piT_eT_bT}
\end{align}
Sequentially using
(\ref{id_iX_star}), (\ref{id_iX_move}), (\ref{id_star_iX}),
(\ref{Lift_PermGen_t_res3}), (\ref{lift_piT_eT_bT}), (\ref{id_star_iX}),
(\ref{id_iX_move}), (\ref{id_iX_star}),
the first term on the right hand side of (\ref{Lift_Lag_T})
becomes
\begin{align*}
F\wedge\starT(\PermDET(\MeT)\wedge\dualvT)
=&
F\wedge \ivT\starT\PermDET(\MeT)
=
-\MeT\wedge \starT\PermDET(\MeT)
=
-(\starT1) i_{\dualMeT}\PermDET(\MeT)
\\
=&
-\tfrac12 (\starT1) \left( i_{\dualMeT}\PermDEO(\piT\MeT) +
i_{\dualMeT}\PermDEO(\piT\MeT) \right)
=
-(\starT1)
i_{\dualMeT}\PermDEO(\MeT)
\\
=&
- \MeT \wedge\starT \PermDEO(\MeT)
=
F \wedge\ivT\starT
\PermDEO(\MeT) = F \wedge\starT(\PermDEO(\MeT)\wedge\dualvT)
\end{align*}

Similarly sequentially using (\ref{id_iX_star}), (\ref{id_iX_move}),
(\ref{id_star_iX}), (\ref{Lift_PermGen_t_res3}), (\ref{lift_piT_eT_bT}),
(\ref{id_star_iX}), (\ref{id_iX_move}), (\ref{id_iX_star}),
(\ref{id_star_pivot}), (\ref{id_star_star}) the second term on the
right hand side of (\ref{Lift_Lag_T}) gives
\begin{align*}
\lefteqn{F\wedge\starT(\PermDBT(\MbT)\wedge\dualvT)} \qquad&
\\
=&
F\wedge \ivT\starT\PermDBT(\MbT)
=
- \MeT\wedge\starT\PermDBT(\MbT)
=
-(\starT1) i_{\dualMeT}\PermDBT(\MbT)
\\
=&
-\tfrac12 (\starT1) \left( i_{\dualMeT}\PermDBO(\piT\MbT)
- i_{\dualMbT}\PermHEO(\piT\MeT) \right)
=
-\tfrac12 (\starT1) \left(i_{\dualMeT}\PermDBO(\MbT) -
i_{\dualMbT}\PermHEO(\MeT) \right)
\\
=&
-\tfrac12 \MeT\wedge\starT\PermDBO(\MbT) + \tfrac12 \MbT \wedge
\starT\PermHEO(\MeT)
=
\tfrac12 F\wedge\ivT \starT\PermDBO(\MbT) -
\tfrac12 \starT F \wedge \ivT\starT\PermHEO(\MeT)
\\
=&
\tfrac12 F\wedge\starT(\PermDBO(\MbT)\wedge\dualvT) -
\tfrac12 \starT(\PermHEO(\MeT)\wedge\dualvT) \wedge \starT F
\\
=&
\tfrac12 F\wedge\starT(\PermDBO(\MbT)\wedge\dualvT) -
\tfrac12 F \wedge \starT \starT(\PermHEO(\MeT)\wedge\dualvT)
\\
=&
\tfrac12 F\wedge\starT(\PermDBO(\MbT)\wedge\dualvT) +
\tfrac12 F \wedge \PermHEO(\MeT)\wedge\dualvT
\end{align*}
It is useful to record from this calculation that
\begin{align}
-\tfrac12 (\starT1) \left(i_{\dualMeT}\PermDBO(\MbT) -
i_{\dualMbT}\PermHEO(\MeT) \right)
=
\tfrac12 F\wedge\starT(\PermDBO(\MbT)\wedge\dualvT) +
\tfrac12 F \wedge \PermHEO(\MeT)\wedge\dualvT
\label{lift_intermediate_step}
\end{align}

Sequentially using (\ref{id_star_star}), (\ref{id_star_iX}),
(\ref{id_star_pivot}), (\ref{id_star_iX}),
(\ref{Lift_PermGen_t_res3}), (\ref{lift_piT_eT_bT}),
(\ref{lift_intermediate_step}) the third term on the right hand side
of (\ref{Lift_Lag_T}) yields
\begin{align*}
\lefteqn{F\wedge\PermHET(\MeT)\wedge\dualvT} \qquad&
\\
=&
-\PermHET(\MeT)\wedge\dualvT\wedge\starT\starT F
=
\PermHET(\MeT)\wedge\starT\MbT
=
\MbT\wedge\starT\PermHET(\MeT)
\\
=&
(\starT1)i_{\dualMbT}\PermHET(\MeT)
=
\tfrac12(\starT1)\left(i_{\dualMbT}\PermHEO(\piT\MeT) -
i_{\dualMeT}\PermDBO(\piT\MbT) \right)
\\
=&
\tfrac12(\starT1)\left(i_{\dualMbT}\PermHEO(\MeT) -
i_{\dualMeT}\PermDBO(\MbT) \right)
\\
=&
\tfrac12 F\wedge\starT(\PermDBO(\MbT)\wedge\dualvT) +
\tfrac12F \wedge \PermHEO(\MeT)\wedge\dualvT
\end{align*}

Finally on sequential use of
(\ref{id_star_star}), (\ref{id_star_iX}), (\ref{id_star_pivot}),
  (\ref{id_star_iX}), (\ref{Lift_PermGen_t_res3}),
(\ref{lift_piT_eT_bT})
\begin{align*}
F\wedge\PermHBT(\MbT)\wedge\dualvT
=&
-\PermHBT(\MbT)\wedge\dualvT\wedge\starT\starT F
=
\PermHBT(\MbT)\wedge\starT\MbT
=
\MbT \wedge\starT \PermHBT(\MbT)
\\
=&
(\starT1)i_{\dualMbT} \PermHBT(\MbT)
=
\tfrac12 (\starT1) \left(
i_{\dualMbT} \PermHBO(\piT\MbT) + i_{\dualMbT}
\PermHBO(\piT\MbT)\right)
\\
=&
(\starT1)i_{\dualMbT} \PermHBO(\MbT)
\end{align*}
and so by reversing this sequence of steps
\begin{align*}
F\wedge\PermHBT(\MbT)\wedge\dualvT
=
F\wedge\PermHBO(\MbT)\wedge\dualvT
\end{align*}

Hence (\ref{Lift_Lag_T}) simplifies to
\begin{equation}
\begin{aligned}
2\cc\LagT =& F\wedge\starT(\PermDEO(\ivT F)\wedge\dualvT) +
F\wedge\starT(\PermDBO(\ivT\starT F)\wedge\dualvT)
\\
&+ F\wedge\PermHEO(\ivT F)\wedge\dualvT +
F\wedge\PermHBO(\ivT\starT F)\wedge\dualvT
\end{aligned}
\label{Lift_Lag_T_res}
\end{equation}
i.e. the constitutive tensors $\zeta_t$ in the action may be replaced by
$\zeta_0$ and hence the metric dependence of $\Lambda_t$ is seen to
reside solely in $\star_t$, $V_t$ and $\tilde V_t$.

The derivative of (\ref{Lift_Lag_T_res}) at $t=0$ is given by
\begin{equation}
\begin{aligned}
\cc\LagD =& F\wedge\starD(\PermDE(\ivF)\wedge\dualv)
+F\wedge\star(\PermDE(\ivD F)\wedge\dualv)
+F\wedge\star(\PermDE(\ivF)\wedge\dualvD)
\\ &
+F\wedge\starD(\PermDB(\ivstarF)\wedge\dualv)
+F\wedge\star(\PermDB(\ivD\star F)\wedge\dualv)
+F\wedge\star(\PermDB(\iv\starD F)\wedge\dualv)
\\ &
+F\wedge\star(\PermDB(\ivstarF)\wedge\dualvD) +F\wedge\PermHE(\ivD
F)\wedge\dualv +F\wedge\PermHE(\ivF)\wedge\dualvD
\\ &
+F\wedge\PermHB(\ivD \star F)\wedge\dualv +F\wedge\PermHB(\iv
\starD F)\wedge\dualv +F\wedge\PermHB(\iv \star F)\wedge\dualvD
\end{aligned}
\label{Stress_LagD_1}
\end{equation}
where  the subscript $0$  is omitted on the right hand side.

To determine the drive forms, observe that there are three
different types of term in (\ref{Stress_LagD_1}) which contain
$\starD$, $\vD$ or $\dualvD$. Since
$\dualvD=(g(V))\dot{}=g(\dot{V})+\gD(V)$, terms in $\starD$, $\vD$
and $\gD(V)$ can be collected to give:
\begin{align}
2\cc\LagD = \LagD_{\starD} + \LagD_{\vD} + \LagD_{\gD(V)}
\label{Stress_LagD_split}
\end{align}
where
\begin{equation}
\begin{aligned}
2\cc\LagD_{\starD} =& F\wedge\starD(\PermDE(\ivF)\wedge\dualv)
+F\wedge\starD(\PermDB(\ivstarF)\wedge\dualv)
\\&
+F\wedge\star(\PermDB(\iv\starD F)\wedge\dualv)
+F\wedge\PermHB(\iv \starD F)\wedge\dualv \,,
\end{aligned}
\label{Stress_LagD_starD}
\end{equation}
\begin{equation}
\begin{aligned}
2\cc\LagD_{\vD} =& F\wedge\star(\PermDE(\ivD F)\wedge\dualv)
+F\wedge\star(\PermDE(\ivF)\wedge g(\vD))
\\&
+F\wedge\star(\PermDB(\ivD\star F)\wedge\dualv)
+F\wedge\star(\PermDB(\ivstarF)\wedge g(\vD))
\\&
+F\wedge\PermHE(\ivD F)\wedge\dualv +F\wedge\PermHE(\ivF)\wedge
g(\vD)
\\&
+F\wedge\PermHB(\ivD \star F)\wedge\dualv +F\wedge\PermHB(\iv
\star F)\wedge g(\vD)
\end{aligned}
\label{Stress_LagD_vD}
\end{equation}
and
\begin{equation}
\begin{aligned}
2\cc\LagD_{\gD(V)} =& F\wedge\star(\PermDE(\ivF)\wedge\gD(V))
+F\wedge\star(\PermDB(\ivstarF)\wedge\gD(V))
\\&
+F\wedge\PermHE(\ivF)\wedge\gD(V) +F\wedge\PermHB(\iv \star
F)\wedge\gD(V)
\end{aligned}
\label{Stress_LagD_gDv}
\end{equation}

The third term on the right hand side of (\ref{Stress_LagD_starD}) may
be expressed as
\begin{align*}
F\wedge\star(\PermDB(\iv\starD F)\wedge\dualv) =&
F\wedge\iv\star(\PermDB(\iv\starD F)) =
-\ivF\wedge\star\PermDB(\iv\starD F)
\\
=& \iv\starD F\wedge\star \PermHE(\ivF) = -\starD F\wedge\iv\star
\PermHE(\ivF)
\\
=& -\iv\star \PermHE(\ivF)\wedge\starD F =
-F\wedge\starD\star(\PermHE(\ivF)\wedge\dualv)
\end{align*}
using sequentially (\ref{id_iX_star}), (\ref{id_iX_move}),
(\ref{Media_Zeta_adjoint}), (\ref{id_iX_move}), (\ref{id_wedge}),
(\ref{id_starT_pivot}).
The fourth term on the right hand side of (\ref{Stress_LagD_starD}) may
be expressed as
\begin{align*}
F\wedge\PermHB(\iv \starD F)\wedge\dualv =&
-F\wedge\dualv\wedge\PermHB(\iv \starD F) =
-F\wedge\dualv\wedge\star\star\PermHB(\iv \starD F)
\\
=& -\star\PermHB(\iv \starD F)\wedge\star(F\wedge\dualv) =
\ivstarF\wedge\star\PermHB(\iv \starD F)
\\
=& \iv \starD F\wedge\star\PermHB(\ivstarF) = -\starD
F\wedge\iv\star\PermHB(\ivstarF)
\\
=& -\iv\star\PermHB(\ivstarF)\wedge\starD F = -F\wedge\starD
\star(\PermHB(\ivstarF)\wedge\dualv)
\end{align*}
using sequentially (\ref{id_wedge}), (\ref{id_star_star}),
(\ref{id_starT_pivot}), (\ref{id_iX_star}),
(\ref{Media_Zeta_adjoint}), (\ref{id_iX_move}), (\ref{id_wedge}),
(\ref{id_starT_pivot}).

Hence from (\ref{Mink_tau_a_Mink})
\begin{equation}
\begin{aligned}
2\cc\LagD_{\starD} =& F\wedge\starD \Big(
\PermDE(\ivF)\wedge\dualv +\PermDB(\ivstarF)\wedge\dualv
-\star(\PermHE(\ivF)\wedge\dualv)
-\star(\PermHB(\ivstarF)\wedge\dualv) \Big)
\\
=& F\wedge\starD G
\\
=& \deltae^a\wedge\big(F \wedge \ia \star G -\ia G \wedge \star
F\big)
\end{aligned}
\label{Stress_tau_starD}
\end{equation}

To collect terms in $\LagD_{\vD}$ observe that by differentiating
(\ref{Lift_V_t}), $\vD=\lambda V$ where
$\lambda=\deltae^a(V)\,V_a$ one has
\begin{align*}
2\cc\LagD_{\vD} =& 2\lambda\big(
F\wedge\star(\PermDE(\ivF)\wedge\dualv)
+F\wedge\star(\PermDB(\ivstarF)\wedge\dualv)
\\&
+F\wedge\PermHE(\ivF)\wedge\dualv +F\wedge\PermHB(\iv \star
F)\wedge\dualv \big)
\\
=& 2 \lambda F\wedge\star G
= 2 \iv\deltae^a\ V_a F\wedge\star G
= 2 \deltae^a\wedge V_a\iv(F\wedge\star G)
\end{align*}

The first two terms on the right hand side
of (\ref{Stress_LagD_gDv}) become
\begin{align*}
\lefteqn{ F\wedge\star(\PermDE(\ivF)\wedge\gD(V))
+F\wedge\star(\PermDB(\ivstarF)\wedge\gD(V)) }\qquad&
\\
=&F \wedge \star (\iv G\wedge \dot{g}(V)) = \iv G\wedge \dot{g}(V)
\wedge \star F = -\dot{g}(V)\wedge \iv G\wedge \star F
\end{align*}
and the last two terms on the right hand side
of (\ref{Stress_LagD_gDv}) become
\begin{align*}
F\wedge\PermHE(\ivF)\wedge\gD(V) +F\wedge\PermHB(\iv \star
F)\wedge\gD(V) = -\dot{g}(V)\wedge F \wedge \iv\star G
\end{align*}
so using  $\gD(V)
= \deltae^a(V)\ e_a + \deltae^a \ e_a(V) = 2\deltae^a \ V_a
+\iv(\deltae_a\wedge e^a)$ one has
\begin{align*}
2\cc\LagD_{\gD(V)} =& -\dot{g}(V)\wedge (\iv G\wedge \star F + F
\wedge \iv\star G)
\\
=& -(2\deltae^a \ V_a +\iv(\deltae_a\wedge e^a))\wedge (\iv
G\wedge \star F + F \wedge \iv\star G)
\\
=& -2\deltae^a\wedge (\iv G\wedge \star F + F \wedge \iv\star G) +
\deltae_a\wedge e^a\wedge \iv(\iv G\wedge \star F + F \wedge
\iv\star G)
\\
=& \deltae\wedge\big( -2(\iv G\wedge \star F + F \wedge \iv\star
G) + e^a\wedge(\ivF \wedge \iv\star G-\iv G\wedge \star F) \big)
\end{align*}
Adding this to $2\LagD_{\vD}$ gives
\begin{align*}
2\cc\LagD_{\vD}+2\cc\LagD_{\gD(V)} =& \deltae^a\wedge\big(
2V_a(\ivF\wedge\star G - \iv G\wedge \star F) + \deltae^a\wedge
e^a\wedge(\ivF \wedge \iv\star G-\iv G\wedge \star F) \big)
\\
=& 2 V_a \deltae^a\wedge(\ivF\wedge\star G - \iv G\wedge \star F)
- \deltae^a\wedge e^a\wedge\iv(\ivF\wedge\star G - \iv G\wedge
\star F)
\end{align*}
Using the relation $\star G=\iv\star G\wedge\dualv+\iv\star\iv G$
and the similar relation for $\star F$, the combination  above may
be writen
\begin{align*}
\lefteqn{\ivF\wedge\star G - \iv G\wedge \star F} \qquad\qquad&
\\
=& \ivF\wedge\ivstarG\wedge\dualv + \ivF\wedge\iv{\star}\ivG -
\ivG\wedge\ivstarF\wedge\dualv - \ivG\wedge\iv{\star}\ivF
\\
=& \star\GenPoyn - \iv(\ivF\wedge\star\ivG - \ivG\wedge\star\ivF)
= \star\GenPoyn
\end{align*}
where the 1-form
\begin{align}
\GenPoyn=\star\left(\ivF\wedge\ivstarG\wedge\dualv +
\ivstarF\wedge\ivG\wedge\dualv\right)
\label{intro_general_poyntin}
\end{align}
Hence
\begin{align}
2\cc\LagD_{\vD}+2\cc\LagD_{\gD(V)} =& \deltae^a\wedge \big(2 V_a
{\star}\GenPoyn - e^a\wedge\iv{\star}\GenPoyn\big)
\label{Stress_tau_dualv}
\end{align}
Adding together (\ref{Stress_tau_starD}) and
(\ref{Stress_tau_dualv}) gives finally:
\begin{align}
2\cc\LagD=\deltae^a\wedge \big(F \wedge \ia \star G -\ia G \wedge
\star F + 2 V_a {\star}\GenPoyn - e^a\wedge\iv{\star}\GenPoyn
\big)
\end{align}
Hence the drive forms are given by
\begin{align}
\cc\tau_a =& \tfrac12 \big(F \wedge \ia \star G -\ia G \wedge
\star F \big) + V_a\star\GenPoyn - \tfrac12
e_a\wedge\iv\star\GenPoyn \label{intro_Abraham_Stress_forms}
\end{align}
with associated  stress-energy-momentum tensor:
\begin{align}
T= \tfrac12 \Big(i_a F\otimes i^a G +  i_a G\otimes i^a F -
\star(F\wedge\star G) g +  \dualv \otimes \GenPoyn + \GenPoyn
\otimes \dualv \Big) \label{intro_Abraham_Stress_T}
\end{align}
The tensor $T$ above  coincides in Minkowski spacetime with that
attributed historically to Abraham. It is derived here in a
considerably wider context.


In terms of comoving fields the drive forms can be written:
\begin{equation}
\begin{aligned}
\cc\tau_{a} =& V_a \iv(\Me \wedge \star\Md + \Mh \wedge \star\Mb)
- \tfrac{1}{2}(\Me \wedge \ia\star\Md + \ia \Md \star\Me )
\\ &
- \tfrac{1}{2}(\Mh \wedge \ia\star\Mb + \ia \Mb \star\Mh )
 + 2v_a \Me \wedge \Mh \wedge \dualv + e_a \wedge \Me \wedge \Mh
\end{aligned}
\label{Abraham_tau_a}
\end{equation}
and hence
\begin{equation}
\begin{aligned}
T =& -\tfrac{1}{2} (\Me \otimes \Md + \Md \otimes \Me)
-\tfrac{1}{2} (\Mh \otimes \Mb + \Mb \otimes \Mh)
\\ &+
\tfrac{1}{2} (
g(\tilde{\Me},\tilde{\Md})
+ g(\tilde{\Mh},\tilde{\Mb})) ( g + 2 \dualv \otimes \dualv ) +
(\dualv \otimes \tilde{S} + \tilde{S} \otimes{V})
\end{aligned}
\label{Abraham_T_ebdh}
\end{equation}
where the Poynting 1-form
\begin{align*}
\tilde{S} = \star( \dualv \wedge \Me \wedge \Mh)
\end{align*}

\def\STAR{\star}
\newcommand{\beq}[1]{{\begin{equation}#1\end{equation}}}

One may express the expressions above in terms of comoving
polarisation $1-$forms $\bfp$ and   magnetisation  $1-$forms
$\bfm$, defined in terms of comoving electromagnetic fields by
\beq{\bfd=\bfe +\bfp} \beq{\bfh=\bfb-\bfm}
 Thus \beq{G=F + \PPi }
where \beq{\PPi =\bfp \wdd \tilde{V} + \star\,(\bfm \wdd
\tilde{V}) } Then one finds
\begin{align*}
{ \tau_c=\tau^{1}_c + \tau^{2}_c + \tau^{3}_c + \tau^{4}_c }
\end{align*}
where
\begin{align*}
2\cc\tau^{1}_c =& i_c \STAR G \wdd F - i_c G \wdd \STAR F \,,
\\
2\cc\tau^{2}_c =& V_c\,\left( \bfp \wdd \STAR F + \bfm \wdd F
-(\bfp \wdd \Mb + \bfm \wdd \Me) \wdd \tilde V \right) \,
\\
2\cc\tau^{3}_c =& -(i_c F) \wdd (\bfm \wdd\tilde V) - (i_c\STAR F)
\wdd \STAR(\bfm \wdd \tilde V)
\\
2\cc\tau^{4}_c =& -V_c\,(\bfp \wdd \STAR F + \bfm \wdd F) -
V_c\,\tilde V \wdd (\bfp\wdd \Mb + \bfm \wdd\Me) - e_c\wdd(\bfp
\wdd \Mb + \bfm\wdd \Me)
\end{align*}

The above are valid for all simple media in arbitrary gravitational
fields.  For a simple medium, which may be inhomogeneous, anisotropic
and intrinsically magneto-electric, at rest in an inertial frame in
{\it Minkowski spacetime} with Minkowski coordinates $\{t, \vec{x}\}$
one has (in Euclidean notation)
\begin{align*}
\dualv= -dt \quad , \quad \Me = \vec{E} \cdot d\vec{x} \quad ,
\quad \Mb = \vec{B} \cdot d\vec{x} \quad , \quad \Mh = \vec{H}
\cdot d\vec{x} \quad , \quad \Md = \vec{D} \cdot d\vec{x}
\end{align*}
and
\begin{align*}
g(\tilde{\Me},\tilde{\Md}) = \vec{E}\cdot \vec{D} \quad , \quad
g(\tilde{\Mh},\tilde{\Mb}) = \vec{H} \cdot \vec{B}
\end{align*}

\begin{align*}
\tilde{S} = - (\vec{E} \times \vec{H}) \cdot d\vec{x}
\end{align*}
The coordinate components of the stress-energy-momentum tensor
follow as
\begin{equation}
\begin{aligned}
T_{00} &= \tfrac{1}{2} (\vec{E}\cdot \vec{D} + \vec{H} \cdot
\vec{B})
\\
T_{ij} &= -\tfrac{1}{2} ( E_i D_j + E_j D_i )
 -\tfrac{1}{2} ( H_i B_j + B_j H_i ) + \tfrac{1}{2} \delta_{ij} (
\vec{E}\cdot \vec{D} + \vec{H} \cdot \vec{B})
\\
T_{0k} &= T_{k0} = -(\vec{E} \times \vec{H})_k
\end{aligned}
\label{Abraham_T_components}
\end{equation}


\section{Conclusions  } \label{concl}
\def\nn{\cal N}

Natural assumptions made above for the dependence of the
constitutive tensor $Z$ on the normalised 4-velocity of a simple
medium have led via a non-trivial  variational argument to a
contribution to the stress-energy-momentum tensor (involving
phenomenological electromagnetic interactions with bulk matter)
that coincides with that suggested by Abraham under more
restricted circumstances. Although natural, the assumptions based
on physical considerations are not, however, necessarily the
simplest to make.

If  $\PermTen$ is chosen to be  independent of the metric and
hence $\tilde V$, with $\PermTenT=\PermTenO$ and  $\PermTenD=0$ so
that $G=Z_0(F)$ in all gravitational fields, one obtains
immediately from the above variational calculations
(\ref{Mink_general_tau_a}) the drive forms
\begin{align}
\cc\tau_a = \tfrac12 \big( F \wedge \ia \star G - \ia G \wedge
\star \,F \big) \label{intro_sym_Minkowski_Stress_forms}
\end{align}
 and the
associated  stress-energy-momentum tensor
\begin{align}
T= \tfrac12 i_a G\otimes i^a F  + \tfrac12 i_a F\otimes i^a G -
\tfrac12 \star(F\wedge\star\, G) g
\end{align}
showing clearly its independence of the 4-velocity of the medium.
It is of interest to note that such a tensor coincides with that
obtained by symmetrising the one proposed by Minkowski.

In the absence of a generally accepted relativistic covariant
description of deformable matter interacting with electromagnetic
fields, the adoption of a particular \SEM tensor for the
electromagnetic field alone in polarisable (and possible
magneto-electric)  media must remain a matter of expediency.
However,  useful models for the total \SEM tensor for such systems
can benefit from the use of sufficiently general phenomenological
descriptions of the electromagnetic properties of moving media
compatible with relativistic covariance. For example a
thermodynamically inert (pressureless, cold) fluid can be modelled
by adding  the electromagnetic stress-energy-momentum tensor
(\ref{intro_Abraham_Stress_T})  to the matter
stress-energy-momentum tensor $\frac{m_0}{c\epsilon_0} \,{\nn}
\widetilde {V} \otimes \widetilde {V} $ where $\nn$ is a scalar
number density field, $m_0$ some constant with the dimensions of
mass and $V$ the unit time-like 4-velocity field of the fluid.
Supplemented with continuity conditions, the vanishing divergence
of such a combination yields the dynamics of the system and  with
prescribed boundary conditions at an interface separating such
media with different properties one may compute bulk forces and
torques.

A review has also been given of the symmetry constraints expected
of the total \SEM tensor particularly when this is considered to
be a source of relativistic gravitation. This led to a definition
in terms of  a variational derivative and a consideration of the
response of the electromagnetic constitutive properties to
gravitational perturbations.  It is suggested that \SEM tensors
parameterised by a self-adjoint constitutive tensor $Z$ offer a
viable means to explore the electromagnetic properties of a range
of inhomogeneous, anisotropic and possibly magneto-electric
continua, at least in regions where dispersion and losses can be
ignored to a first approximation. This formulation suggests a
method to determine the properties of $Z$ by exploring its
phenomenological response to electromagnetic fields in arbitrarily
moving reference frames and variable gravitational fields. It
opens up the possibility of performing such experiments in new
environments such as those carried out under terrestrial free-fall
or space station situations or in astrophysical contexts.

\vspace{0.5cm} \noindent {\bf Acknowledgements} The authors are
grateful D. Burton and A. Noble for helpful discussions and to the
EPSRC and  Framework 6 (FP6-2003-NEST-A) for financial support for
this research.

\vfill\eject

\section{Appendix } \label{appendix}

\def\DOT{\dot{\,\,}} Using the notation established in the text, this
appendix derives the useful formula (\ref{Lift_starD}) relating
$(\star \Psi)\DOT{}$ to $\star \dot\Psi$ where $\Psi\in
\GamLamM{p}$. Let $I$ denote a multi-index constructed from the
single indices $a,b,c\ldots $ in the range $0,1,2,3$ where the
components of the metric tensor $g$ and $\Psi$ in an
$g$-orthonormal basis $\{e^c\}$ are respectively $\eta_{ab}$ and
$\Psi_I$. Thus

\begin{align*}
\Psi=&\Psi_I \, e^I
\end{align*}
and
\begin{align}
\dot\Psi=&\dot\Psi_I\,e^I +
\Psi_I\,{(e^I)}\DOT\label{A1}
\end{align}
Since $e^I$ is the exterior product of $p$ 1-forms
\begin{align*}
(e^I)\DOT=\dot e^c\wedge i_c(e^I)
\end{align*}
Similarly, since the basis is orthonormal
\begin{align*}
(\star e^I)\DOT=\dot e^c\wedge i_c(\star e^I)
\end{align*}
Thus using (\ref{A1})
\begin{align*}
\dot\Psi_I e^I=\dot\Psi - \dot e^c\wedge i_c\Psi
\end{align*}
Applying $\star$ to this gives
\begin{align}
\dot\Psi_I \star e^I=\star\dot\Psi - \star(\dot e^c\wedge
i_c\Psi)\label{A2}
\end{align}
But
\begin{align*}
(\star\Psi)\DOT=&(\Psi_I\star e^I)\DOT\\=&\dot\Psi_I\star e^I +
\Psi_I(\star e^I)\DOT\\=& \dot\Psi_I\,\star e^I +  \Psi_I\,\dot
e^c\wedge i_c(\star e^I)\\=&\dot\Psi_I\,\star e^I + \dot e^c
\wedge i_c(\star \Psi)
\end{align*}
Substituting from (\ref{A2}) yields the relation
\begin{align*}
(\star \Psi)\DOT= \dot e^c \wedge i_c(\star \Psi) - \star(\dot e^c
\wedge i_c\Psi) + \star \dot\Psi
\end{align*}



\end{document}